\documentclass[pre,aps,10pt,superscriptaddress,twocolumn,floatfix]{revtex4-1}

\usepackage[nice]{nicefrac}
\usepackage{graphicx,color}
\usepackage{amsmath,amssymb,bm,bbm}
\usepackage{amsmath}
\usepackage{braket}
\usepackage{float}
\usepackage{placeins}

\usepackage[plainpages=false,pdfpagelabels,colorlinks=true,linkcolor=red,urlcolor=blue,citecolor=blue,pdftitle={},pdfauthor={},pdfdisplaydoctitle=true,pdfduplex=DuplexFlipLongEdge]{hyperref}

\definecolor{darkred}{rgb}{0.90,0.2,0.2}
\definecolor{darkgreen}{rgb}{0,0.60,.2}
\definecolor{darkblue}{rgb}{0.1,0.3,1}
\definecolor{grey}{cmyk}{0,0,0,0.25}
\definecolor{orange}{cmyk}{0,0.6,0.8,0}

\begin{document}

\title{Local integrals of motion in dipole-conserving models with Hilbert space fragmentation}

\author{Patrycja  \L yd\.{z}ba}
\affiliation{Institute of Theoretical Physics, Wroclaw University of Science and Technology, 50-370 Wroc{\l}aw, Poland}
\author{Peter Prelovšek}
\affiliation{Department of Theoretical Physics, J. Stefan Institute, SI-1000 Ljubljana, Slovenia}
\author{Marcin Mierzejewski}
\affiliation{Institute of Theoretical Physics, Wroclaw University of Science and Technology, 50-370 Wroc{\l}aw, Poland}

\begin{abstract}
Hilbert space fragmentation is an ergodicity breaking phenomenon, in which Hamiltonian shatters into exponentially many dynamically disconnected sectors. In many fragmented systems, these sectors can be labelled by statistically localized integrals of motion, which are {\em nonlocal} operators. We study the paradigmatic nearest-neighbor pair hopping (PH) model exhibiting the so-called strong fragmentation. We show that this model hosts {\em local} integrals of motion (LIOMs), which correspond to frozen density modes with long wavelengths. The latter modes become subdiffusive when longer-range pair hoppings are allowed. Finally, we make a connection with a tilted (Stark) chain. Contrary to the dipole-conserving effective models, the tilted chain is shown to support either Hamiltonian or dipole moment as an LIOM. Numerical results are obtained from a numerical algorithm, in which finding LIOMs is reduced to the data compression problem.
\end{abstract}
\maketitle

\textit{Introduction.}---Over the last two decades, a considerable effort has been made to understand whether an isolated quantum system thermalizes after being driven out of equilibrium. The thermal state is  determined by just a few  local integrals of motion (LIOMs) usually corresponding to conservations of energy and particle number.
Moreover, their long-wavelength excitations, i.e., energy and density modes, attenuate according to the Fick's law of diffusion~\cite{Bertini_2021}.
It has been confirmed, also experimentally~\cite{Trotzky_2012,Kaufman_2016,Clos_2016}, that interacting systems typically thermalize. This is mostly understood in terms of the eigenstate thermalization 
hypothesis~(ETH)~\cite{Srednicki_1994,Deutsch_2018,D_Alessio_2016}.

It is natural to look for interacting systems that fail to thermalize. The most studied examples are integrable systems~\cite{PhysRevX.8.021030,Kinoshita2006,rigol_2011,Franchini_2017}, which have {\em an extensive number} of LIOMs.
These local (or quasilocal~\cite{PhysRevLett.106.217206}) conserved operators affect the dynamics of other local observables. Specifically, the steady-state expectation values of local observables follow the predictions of the generalized Gibbs ensemble~\cite{Rigol_2007,Vidmar_2016,Langen_2015,PhysRevB.87.064201,PhysRevLett.131.060401}, which is set by the LIOMs. Additionally, LIOMs pose limits on the decay (in time) of correlation functions via the Mazur bound~\cite{SUZUKI1971277,MAZUR1969533,Zotos}. The inverse relation also holds, and the non-vanishing correlations at infinite time imply the existence of local or quasilocal integrals of motion ~\cite{mierzejewski2020}.

More exotic violations of the ETH are also studied, such as the many-body localization (MBL). This phenomenon may arise from the intricate interplay of disorder and interactions that leads to emergent LIOMs, known as $l$-bits~\cite{Nandkishore_2015,RevModPhys.91.021001}. We note that the existence of strict MBL in macroscopic systems is currently a subject of debate~\cite{Suntajs_2020,Suntajs_2020b,LeBlond_2021,Krajewski_2023,prelovsek23}. Similar physics has been proposed for Stark systems~\cite{van_Nieuwenburg_2019,Doggen_2021,Yao_2021,Buca_2022,Zisling_2022}, in which the role of disorder is taken over by tilted potential. In the large tilt limit, its non-equilibrium dynamics is well captured by the~effective models that strictly
conserve the dipole moment~\cite{PhysRevLett.130.010201,Scherg_2021,Moudgalya_2021}. The simplest one is the PH model that exhibits the Hilbert space fragmentation~\cite{Pai_2019,Khemani_2020,Moudgalya_2022,Francica_2023,Brighi_2023,will2023realization,Sala_2020}, since its
Hamiltonian shatters into exponentially many blocks (Krylov subspaces) in the site occupation basis. We emphasize that an important step towards understanding this phenomenon is the introduction of statistically localized integrals of motion~\cite{Rakovszky_2020,Moudgalya_2022b}, which label the Krylov subspaces. Nevertheless, these are highly nonlocal operators
and in general it is not obvious how nonlocal conserved operators affect the dynamics of local observables.

Since the presence of restrictions on the asymptotic dynamics of local observables is commonly understood as a manifestation of LIOMs, in this Letter we look for a~connection between the Hilbert space fragmentation and their existence. First, we propose a numerical algorithm that establishes all LIOMs linear in a given
set of operators. We employ it to demonstrate that the long wavelength density modes in the PH model are frozen
and become strict LIOMs in the thermodynamic limit. Next, we argue that these density modes become subdiffusive after incorporating longer-range pair hoppings to the PH model, which break the strong fragmentation. Finally, we make a connection with the full
Stark model. We demonstrate that although both energy and dipole moment are conserved in the
thermodynamic limit~\cite{Nandy_2023}, they correspond to a single LIOM in this case.

\textit{Method.}---We first develop a simple algorithm that determines whether and how many LIOMs can be constructed from a fixed set of operators.  We consider a~Hilbert space of dimension 
$Z$, which is spanned by energy eigenstates, $H|n\rangle=E_n |n\rangle$, and denote matrix elements of observables as $A_{mn}= \langle m|A|n\rangle$. We are interested in Hermitian operators, for which the Hilbert-Schmidt (HS) product is
\begin{equation}
\langle A B \rangle =\frac{1}{Z} \; {\rm Tr}(AB)=\frac{1}{Z}\sum_{m,n} A_{mn} B^{*}_{mn} \label{hs}
\end{equation}
and the HS norm is $ ||A||^2 =\langle A A \rangle =1/Z\sum_{m,n} |A_{mn}|^2$.

Considering an arbitrary set of orthonormal operators, $\langle A^{i}  A^{i'}\rangle =\delta_{i i'}$, we construct the orthogonal transformation
\begin{equation}
\label{eqtrans}
B^{\beta}=\sum_{i} \mathcal{V}_{i\beta} A^{i}\;,
\end{equation}
so that the set $\{B^{\beta}\}$ includes all LIOMs linear in $A^{i}$. If the set $\{A^i \}$ contains only local observables then all generated $B^{\beta}$ are local by construction. 
We note that local observables are the ones that can be written as sums of operators involving a~finite number of sites, i.e., having a finite support.
From now on, we explicitly distinguish LIOMs from other operators using the symbol $Q^{\beta}$ (${\cal B}^{\beta}$) for $B^{\beta}$ that do (do not) commute with the Hamiltonian. Moreover, we work in the Hilbert space with a fixed particle number $N$. Since $N$ is trivially conserved, we do not discuss it or explicitly include in the set of LIOMs.

In order to find LIOMs, we use the infinite time averaging
\begin{equation}
    \overline{A}=\lim_{\tau\rightarrow\infty} \frac{1}{\tau}\int_{0}^{\tau}e^{iHt}Ae^{-iHt}\;dt = \sum_{\substack{m,n\\E_n=E_m}} A_{nm} |n\rangle\langle m|\;.
    \label{ta}
\end{equation}
It is evident that the time averaging does not modify LIOMs ($\bar{Q}^{\beta}=Q^{\beta}$), while it eliminates some of the matrix elements of ${\cal B}^{\beta}$ (${\bar{\cal B}}^{\beta}_{mn}=0$ for $E_m \ne E_n $). Therefore, LIOMs can be singled out just by examining the norms of the time-averaged operators, $||\bar{Q}^{\beta}||^2=1$ whereas $||\bar{{\cal B}}^{\beta}||^2< 1$. 


We note that the nonergodic behavior of the observables $A^i$ is encoded in the matrix elements of $\bar{A}^i$. We store all matrix elements of all $\bar{A}^i$ in 
a single matrix $\mathcal{R}$ so that its $i$-th column contains $A^{i}_{mn}$ for $E_m=E_n$. In Supplemental Material~\cite{supmat}, we demonstrate that
LIOMs can be established from the compression of data stored in $\mathcal{R}$. Such compression can be achieved via singular value decomposition, \mbox{$\mathcal{R}=\mathcal{U} \Sigma \mathcal{V}^{\rm T}$}, where the diagonal matrix \mbox{$\Sigma=\mathrm{diag}(\tilde{\lambda}_1,\tilde{\lambda}_2,...)$} stores the singular values and the orthogonal matrix $\mathcal{V}$ is the same as in Eq.~(\ref{eqtrans}). 
According to the Eckart–Young–Mirsky theorem, the compression amounts to the approximation 
\mbox{$\mathcal{R} \simeq \mathcal{R^{\parallel}}=\mathcal{U} \Sigma^{\parallel} \mathcal{V}^{\rm T}$}, where $\Sigma^{\parallel}$ contains only the largest $\tilde{\lambda}_{\beta}$.
We consider positive $\tilde{\lambda}_{\beta}$ sorted in descending order, and introduce 
notation $\lambda_{\beta}=\tilde{\lambda}^2_{\beta}/Z$. In Ref.~\cite{supmat}, we show 
that the HS norms of the time-averaged operators from Eq.~(\ref{eqtrans}) read
$||\bar{B}^{\beta}||^2=\lambda_{\beta} \le 1$. Based on the arguments from the preceding 
paragraph, we note that the largest $\lambda_{\beta}=1$ define LIOMS, i.e., 
$Q^{\beta}=\bar{B}^{\beta}=B^{\beta}$.



\textit{Pair hopping model.}---We use the introduced method to investigate LIOMs in the simplest effective model for the~tilted chain. Namely, we consider the PH model with $L$ sites and $N=L/2$  spinless fermions~\cite{Moudgalya_2021},
\begin{equation}
\label{eqH1}
    H_1=\sum_{i=1}^{L-3} (c^\dagger_{i}c^\dagger_{i+3}c_{i+2}c_{i+1} + \text{H.c.})\;.
\end{equation}
Here, $c^\dagger_{i}$ ($c_{i}$) creates (annihilates) a spinless fermion at site~$i$, and we define a traceless site occupation as $n_{i}=c^\dagger_{i}c_{i}-N/L$. This~model conserves the dipole moment $M=\sum_{i} i n_{i}$ as well as the~sublattice particle numbers $n_\text{even}=\sum_{i\in\text{even}} n_{i}$ and $n_\text{odd}=N-n_\text{even}$. 

The PH model manifests the strong fragmentation, so that the~size of the largest Krylov subspace is exponentially smaller than the dimension of the~Hilbert space~$Z$~\cite{Moudgalya_2021,PhysRevB.103.134207}. While the existence of blocks appears to affect the non-equilibrium dynamics leading to the lack of thermalization~\cite{Sala_2020}, it remains unclear whether it is linked to the existence of local (or quasilocal) LIOMs~\cite{Calabrese_2016,mierzejewski2020}. Recall that the latter are different than the statistically localized integrals of motion introduced in~\cite{Rakovszky_2020,Moudgalya_2022b}. Moreover, the density modes are expected to undergo a~subdiffusive relaxation, in agreement with the~fracton hydrodynamics~\cite{Nandkishore_2019,Gromov_2020,PhysRevE.107.034142,Grosvenor_2022}, only after longer-range pair hoppings are included in the model (see~\cite{Feldmeier_2020,PhysRevB.101.214205} for results in classical circuit models). Below, we demonstrate that establishing LIOMs helps to understand these features.

It has been argued in Ref.~\cite{Sala_2020} that the density-density correlations $\langle n_i(t) n_i\rangle$ do not decay to zero at long times in the PH model. Therefore, it is straightforward to look for LIOMs that are linear combinations of $n_{i}$. We emphasize that in the subspace with a fixed number of particles, $n_{i}$ are neither independent ($\sum_{i=1}^{L} n_{i}=0$) nor orthogonal ($\langle n_i n_j \rangle \neq \delta_{ij}$). 
Hence, we select the set of independent occupations $\{n_i \}$ with $i\le L-1$, which we then orthogonalize,
$\sum_{j,k=1}^{L-1} U_{aj} \langle n_{j} n_{k} \rangle U^{T}_{kb}=\delta_{ab} \sigma_a $, to obtain 
the set  ${\cal N}$ of orthonormal operators $A^a=\sum_{i=1}^{L-1} \frac{1}{\sqrt{\sigma_a}} U_{ia} n_i$~\cite{supmat}. Nevertheless, one can still express $B^\beta$ from Eq.~(\ref{eqtrans}) as linear combinations of occupations, $B^\beta=\sum_{a=1}^{L-1} \mathcal{V}_{a\beta} A^{a} = \sum_{a,i=1}^{L-1} \frac{1}{\sqrt{\sigma_a}} \mathcal{V}_{a\beta} U_{ia} n_i =\sum_{i=1}^{L-1} \tilde{\mathcal{V}}_{i\beta} n_i$.


\begin{figure}[t!]
\centering
\includegraphics[width=\columnwidth]{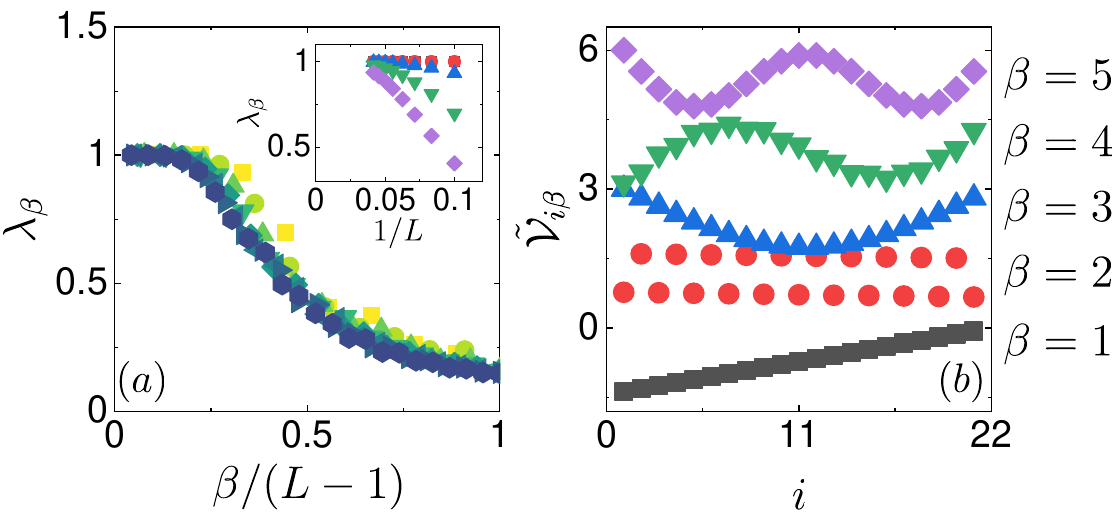}
\caption{$\overline{B}^\beta$ established from $A^i \in {\cal N}$ for the~PH model from Eq.~(\ref{eqH1}). (a) The HS norms $\lambda_\beta$ as functions of $\beta/(L-1)$. We consider $L=10,...,24$ and darker colors represent larger systems. The inset shows the~finite size scaling of the five largest $\lambda_\beta$. (b) Components of the rotated operator, $\tilde{\mathcal{V}}_{i\beta}$, for the five largest $\lambda_\beta$ and $L=22$. Colors are consistent with the inset in (a). For clarity, we shift the curves for various $\beta$ along the vertical axis, i.e., $\tilde{\mathcal{V}}_{i\beta}\rightarrow \tilde{\mathcal{V}}_{i\beta}+1.5(\beta-1)$.}
\label{fig1}
\end{figure}

Results for $\lambda_\beta$ are shown in Fig.~\ref{fig1}(a). Simultaneously in Fig.~\ref{fig1}(b), we plot the coefficients $\tilde{\mathcal{V}}_{i\beta}$ shifted along the vertical axis for clarity, i.e., $\tilde{\mathcal{V}}_{i\beta}\rightarrow \tilde{\mathcal{V}}_{i\beta}+1.5(\beta-1)$. 
We find that $\lambda_{1}=\lambda_{2}=1$  for all system sizes, and that they correspond to the dipole moment $M$ (see black squares in Fig.~\ref{fig1}(b)) and the sublattice particle number $n_\text{even}$ (see red circles in Fig.~\ref{fig1}(b)). The other $\lambda_\beta$ within the plateau in Fig.~\ref{fig1}(a), i.e.,  for $ \beta \ge 3$,  appear to increase towards one with a system size, as demonstrated in the inset of Fig.~\ref{fig1}(a). Consequently, there are infinitely many  many LIOMs in the thermodynamic limit. Although we are not able to determine from Fig.~\ref{fig1}(a) whether their number is linear or sublinear in $L$, their presence explains the previously reported lack of thermalization~(see~\cite{MAZUR1969533,SUZUKI1971277,Zotos}). We therefore expect that the conventional measures of quantum chaos will be similar to the ones of integrable systems~(see~\cite{fremling2022exact,Kudo_2003,Gubin_2012}). For the example of spectral statistics see Supplemental Material~\cite{supmat}.


It is worth to highlight that $Q^\beta$ within the plateau in Fig.~\ref{fig1}(a) with $\beta \ge 3$ correspond to density modes with wave vectors $q=\frac{(\beta-2)\pi}{L}$, which are slightly distorted to ensure orthogonality.
The HS norm $||\bar{Q}^\beta||=1$
means that $Q^\beta$ does not have any off-diagonal matrix elements in the energy basis, so that it cannot show any dynamics at any time scale. 
Therefore, these density modes are strictly frozen in the thermodynamic limit.
In the following, we show that longer-range pair hoppings added to $H_1$ unfreeze the density modes and restore the fracton hydrodynamics.


\textit{Extended pair hopping model.}---It has been demonstrated that including longer-range pair hoppings in the~PH model changes the fragmentation from strong to weak, so that the dimension of the~largest block becomes a finite fraction of the total number of states~$Z$~\cite{Khemani_2020}. The~resulting Hamiltonian supports the~weak ETH, in which the majority of eigenstates is thermal but nonthermal outliers (dubbed many-body scars) are allowed~\cite{Moudgalya_2022}. 

It is reasonable to expect that the number of LIOMs decreases when the fragmentation becomes weak. We verify this expectation by studying the extended pair hopping (EPH) model with two additional terms
\begin{equation}
\begin{split}
\label{eqH2}
    H_2= & H_1+\sum_{i=1}^{L-4} (c^\dagger_{i}c^\dagger_{i+4}c_{i+3}c_{i+1} + \text{h.c.})\\
    & + \sum_{i=1}^{L-5} (c^\dagger_{i}c^\dagger_{i+5}c_{i+4}c_{i+1} + \text{h.c.})\;.
\end{split}
\end{equation}
We stress that $H_2$ in Eq.~$(\ref{eqH2})$ no longer commutes with the~sublattice particle numbers $n_\text{even}$ or $n_\text{odd}$.
The same Hamiltonian but without the last term is studied in the Supplemental Material~\cite{supmat}. It yields similar results to those discussed below, however, the finite-size scaling does not provide a clear picture of the thermodynamic limit. In the Supplemental Material \cite{supmat}, we also provide the~derivation starting from the tilted chain that generates all dipole-conserving terms, as the ones from Eq.~(\ref{eqH2}). Nevertheless, the actual parameters in Eq.~(\ref{eqH2}) are settled to one and, so, are not meant to represent a~realistic effective model of the tilted chain.

\begin{figure}[t!]
\centering
\includegraphics[width=\columnwidth]{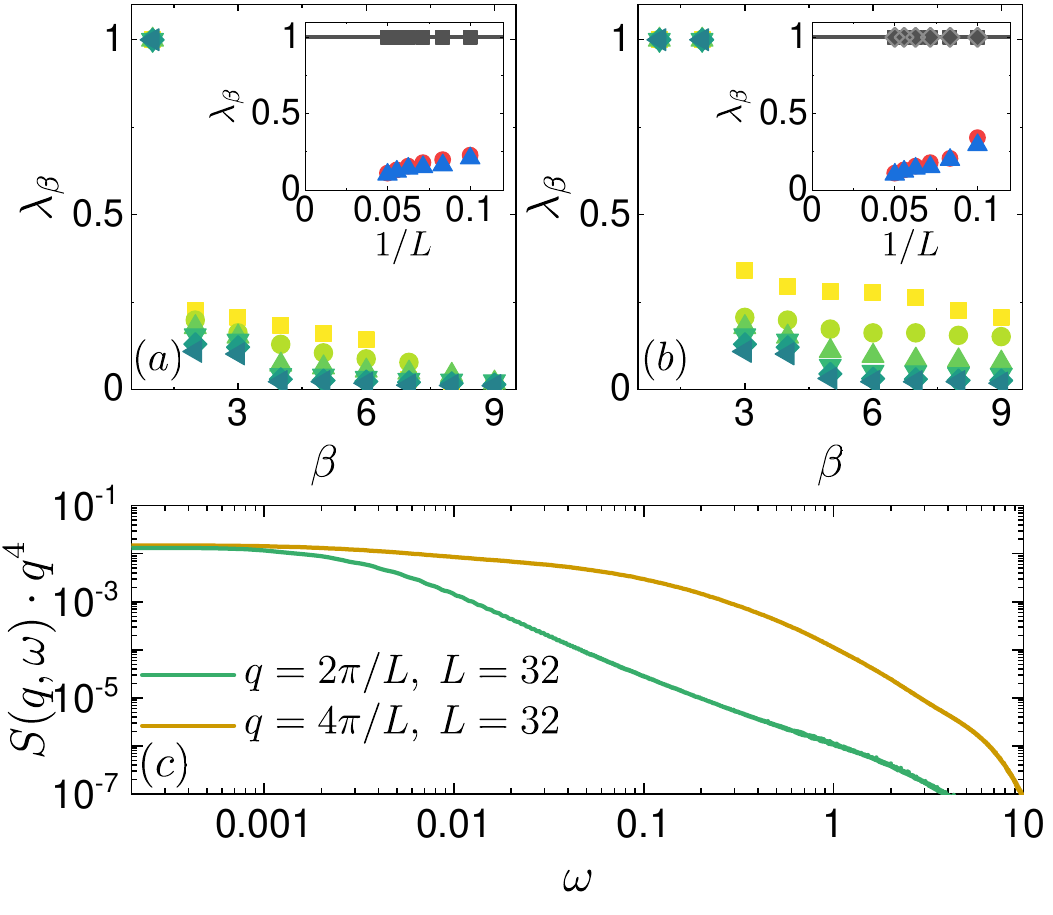}
\caption{(a,b) The same as in Fig.~\ref{fig1}(a) but for the EPH model from Eq.~(\ref{eqH2}), $\beta\le 9$ and
$L=10,...,20$. Results were obtained for various sets of operators: (a) $A^i \in {\cal N}$, (b) $A^i \in {\cal N}_{2E}$. (c)~The dynamical structure factor $S(q,\omega)$ for two smallest $q=2\pi/L, 4\pi/L$.}
\label{fig2}
\end{figure}

In Fig.~\ref{fig2}(a), we show the HS norms $\lambda_ \beta$, where ${B}^\beta$ are linear combinations of occupations, $A^i\in {\cal N}$. It is clear from the inset of Fig.~\ref{fig2}(a) that only $\lambda_1$ is equal to one for all system sizes. It corresponds to the dipole moment $M$ (not shown). Simultaneously, all other $\lambda_\beta$   decrease to zero with a system size. They correspond to the same density modes as shown in Fig.~\ref{fig1}(b). 

The set of operators that we have considered so far, ${\cal N}$, does not allow to construct the Hamiltonian as an~independent LIOM orthogonal to other LIOMs. Therefore, we also study an extended set, ${\cal N}_{2E}$,
which includes all operators from ${\cal N}$ as well as the pair hopping terms,
$c^\dagger_{i}c^\dagger_{i+3}c_{i+2}c_{i+1}+\text{H.c.}$, $c^\dagger_{i}c^\dagger_{i+4}c_{i+3}c_{i+1}+\text{H.c.}$ and $c^\dagger_{i}c^\dagger_{i+5}c_{i+4}c_{i+1}+\text{H.c.}$ Note that $i$ runs through all values for which the site indexes do not exceed $L$. Numerical results obtained for $A^i \in {\cal N}_{2E}$ confirm that the Hamiltonian from Eq.~(\ref{eqH2}) supports only two independent conservation laws, as visible in Fig.~\ref{fig2}(b). The first LIOM is the dipole moment~$M$, while the second LIOM is the~Hamiltonian~$H_2$. 


We now confirm that the density modes relax subdiffusively within the EPH model. Specifically, we numerically calculate the dynamical structure factor  $S(q,\omega)$,
i.e., the dynamical correlation function for density modulations $n_q=1/\sqrt{L} \sum_{i} \cos(q(i-L/2))n_i$ in the~infinite-temperature limit, employing the microcanonical Lanczos method \cite{long03,prelovsek13}. More details can be found in~\cite{supmat}. In the hydrodynamic regime $q \ll 1$, the density modulations should exhibit a slow decay with a characteristic rate $\Gamma_q$, so that the low $\omega \ll 1$ correlations should behave as  $ \pi S(q,\omega)\sim \chi^0 \Gamma_q/(\omega^2+\Gamma_q^2)$ with the corresponding susceptibility $\chi^0  = (1/2) \bar n (1- \bar n)$ and the average density $\bar n = 1/2$. We plot results for two smallest $q=2\pi/L, 4\pi/L$ in Fig.~\ref{fig2}(c), and we find that the relaxation rates scale as $\Gamma_q\propto q^4$, as required for the subdiffusion.

\textit{Stark model.}---Since the Hamiltonians in Eqs.~(\ref{eqH1}) and~(\ref{eqH2}) arise 
as effective models for the strongly tilted chain \cite{supmat,Moudgalya_2021}, 
\begin{equation}
\label{eqH3}
    H_3=t\sum_{i=1}^{L-1}(c_{i}^\dagger c_{i+1}+\text{h.c.}) + FM + V\sum_{i=1}^{L-1} n_i n_{i+1}\;,
\end{equation}
it is instructive to establish LIOMs also in the latter system. We show that the set of LIOMs in the~Stark model differs from the two cases considered earlier. Note that $F$ is the strength of the tilt. Throughout the paper, we fix the hopping integral and the interaction strength to $t=1$ and $V=2$, respectively. For convenience, we denote the translationally invariant part of the Hamiltonian as $H_{3}^{(0)}=H_3-FM$.

\begin{figure}[t!]
\centering
\includegraphics[width=\columnwidth]{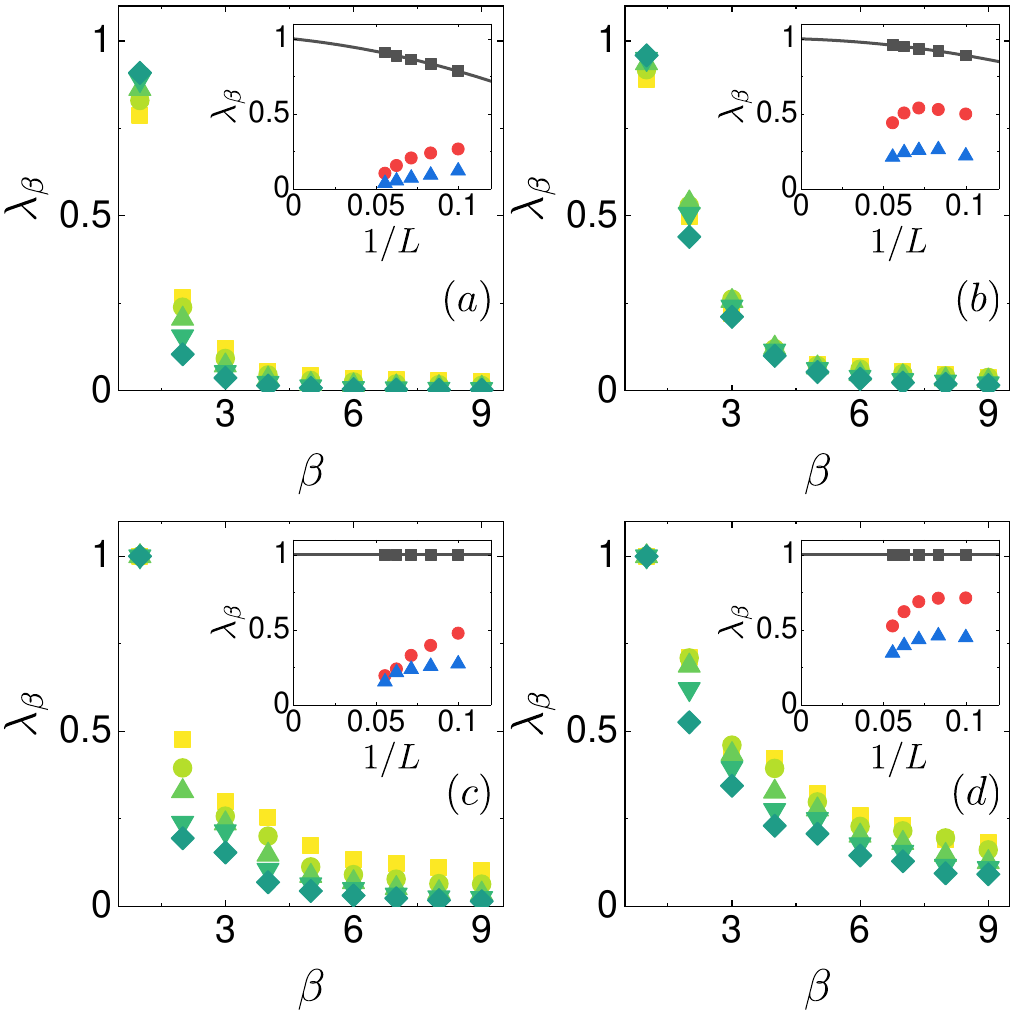}
\caption{
The same as in Figs.~\ref{fig2}(a) and~\ref{fig2}(b) but for the~Stark chain from Eq.~(\ref{eqH3}), $\beta\le 9$ and $L=10,...,18$. Results in different columns were obtained for different fields: (a,c)~$F=1.0$, (b,d)~$F=1.5$. Results in different rows were obtained for different sets of operators: (a,b)~$A^i\in {\cal N}$, (c,d)~$A^i\in {\cal N}_{3E}$.
The solid lines are the second-order polynomial fits to data, serving as guides for the eye.}
\label{fig3}
\end{figure}

The nonequilibrium dynamics of the Stark model has been previously explained in terms of the (approximate) Hilbert space shattering, and the many-body localization in the thermodynamic limit has been proposed~\cite{Doggen_2021,Zisling_2022}. Nevertheless, its transport properties are captured by the~PH model only in small systems and large fields~\cite{Moudgalya_2021}. Moreover, it has been recently argued that the dipole moment $M$ is conserved in the thermodynamic limit and the~density profiles undergo subdiffusive dynamics~\cite{Nandy_2023} (see also~\cite{PhysRevX.10.011042,PhysRevB.108.134304}). This is consistent with our numerical results. In Fig.~\ref{fig3}(a) and \ref{fig3}(b), we demonstrate the HS norm $\lambda_ \beta$ for ${B}^\beta$ that are linear in $A^i\in {\cal N}$ for fields $F=1.0$ and $1.5$, respectively. Only $\lambda_1$ increases towards one with a system size, and it corresponds to the dipole moment $M$. The other $\lambda_\beta$ with $\beta\ge 2$ decrease for $L\ge L^{*}$. They correspond to the density modes from Fig.~\ref{fig1}(b). We find that $L^{*}$ depends on $F$, so that it exceeds the maximal system size available in our numerical simulations for $F\gtrsim 2$. This causes the density modes to appear frozen in finite systems.

We note that the derivation of the dipole moment conservation from~\cite{Nandy_2023} is based on the scaling of the~HS norms $||H_3^{(0)}||^2\propto L$ and $||M||^2\propto L^3$ together with the~orthogonality $\langle H_{3}^{(0)}M\rangle=0$. This brings into question, whether the dipole moment $M$ gives rise to an~LIOM independent of $H_3$ or it simply becomes $H_3$ in the thermodynamic limit. The mere fact that $M$ and $H_3$ are conserved does not imply that their difference  $H_{3}^{(0)}=H_3-FM$ is also an LIOM. Moreover, the vanishing ratio of norms $||\bar{H}_{3}^{(0)}||/||H_3||\le {\cal O}(1/L)$ does not exclude that $H_{3}^{(0)}$ is an LIOM. We emphasize that conserved quantities restrict the relaxation of observables via the Mazur bound \cite{MAZUR1969533,SUZUKI1971277,Zotos}, which involves {\em normalized} LIOMs. Therefore, the significance of $H_{3}^{(0)}$ for the relaxation of observables depends on the ratio $||\bar{H}_{3}^{(0)}||/||H_{3}^{(0)}||$. 

 For this reason, we complement the study by looking for $Q^\beta$ that are linear in $A^i\in {\cal N}_{3E}$, where 
 the set ${\cal N}_{3E}$ includes all operators from ${\cal N}$ as well as the nearest-neighbor interactions $n_i n_{i+1}$ and  hoppings $c_{i+1}^\dagger c_{\i}+\text{h.c.}$ for $i=1,...,L-1$. Note that every term in the Hamiltonian from Eq.~(\ref{eqH3}) belongs to ${\cal N}_{3E}$. 
 Figures~\ref{fig3}(c) and \ref{fig3}(d) reveal an important difference between the Stark chain and the effective model, shown in Fig.~\ref{fig2}(b). In the latter case, the Hamiltonian and dipole moment are two orthogonal LIOMs. In contrast to this, both form only a single LIOM in the Stark chain (in addition to the particle number conservation). When extending the sets of operators from 
${\cal N}$ to ${\cal N}_{3E}$, the dipole moment~$M$ is replaced by the~Hamiltonian $H_3$. Therefore, the conservation of dipole moment is not independent from the conservation of energy. 


Identification of the conservation laws is the starting point for constructing the relevant hydrodynamics. Our results indicate that in the case of the~EPH model one should account for the conservation of the particle number, dipole moment  and energy  \cite{Gromov_2020,PhysRevE.107.034142}. On the other hand, for the tilted Hamiltonian $H_3$ one should use either the energy conservation or the dipole moment conservation \cite{Nandy_2023,PhysRevX.10.011042}, as these two conservation laws are equivalent in the thermodynamic limit.

\textit{Summary.}---In this Letter, we have studied the PH model, which is a paradigmatic model of the Hilbert space fragmentation, and arises in the Schrieffer-Wolff transformation of the Stark model. First, we have put forward a numerical algorithm based on the data compression problem, which generates all LIOMs linear in a given set of operators. Next, we have established that the PH model  hosts an infinite number number of LIOMs in the thermodynamic limit. While the PH model is translationally invariant, the obtained LIOMs are not.  
They correspond to frozen density modes excluding any particle hydrodynamics. On the other hand, density modes decay  when longer-range pair hoppings are allowed and become subdiffusive in agreement with the fracton hydrodynamics. 

We have also revealed an important difference between the Stark chain and its effective models. In the latter cases, the Hamiltonian and dipole moment are two orthogonal LIOMs. In contrast to this, they form only one LIOM in the tilted (Stark) chain.

\acknowledgements 
We acknowledge discussions with J.~Herbrych, A.~G{\l}{\'o}dkowski and L.~Vidmar. We also acknowledge the~support of the National Science Centre, Poland via projects 2023/07/X/ST3/01707 (P.~{\L}.) and 2020/37/B/ST3/00020 (M.~M.) as well as the support of the Slovenian Research Agency via the program P1-0044 (P.~P). Numerical studies in this work have been partially carried out using resources provided by the~Wroclaw Centre for Networking and Supercomputing, Grant No. 579 (P.~{\L}.).

\bibliographystyle{biblev1}
\bibliography{references}

\newpage
\phantom{a}
\newpage
\setcounter{figure}{0}
\setcounter{equation}{0}
\setcounter{table}{0}

\renewcommand{\thetable}{S\arabic{table}}
\renewcommand{\thefigure}{S\arabic{figure}}
\renewcommand{\theequation}{S\arabic{equation}}
\renewcommand{\thepage}{S\arabic{page}}

\renewcommand{\thesection}{S\arabic{section}}

\onecolumngrid

\begin{center}

{\large \bf Supplemental Material:\\
Local integrals of motion in dipole-conserving models with Hilbert space fragmentation}\\

\vspace{0.3cm}

\setcounter{page}{1}

\author{Patrycja  \L yd\.{z}ba}
\affiliation{Institute of Theoretical Physics, Wroclaw University of Science and Technology, 50-370 Wroc{\l}aw, Poland}
\author{Peter Prelovšek}
\affiliation{Department of Theoretical Physics, J. Stefan Institute, SI-1000 Ljubljana, Slovenia}
\author{Marcin Mierzejewski}
\affiliation{Institute of Theoretical Physics, Wroclaw University of Science and Technology, 50-370 Wroc{\l}aw, Poland}

\end{center}

\vspace{0.6cm}

\twocolumngrid

\label{pagesupp}

\section{Numerical algorithm determining LIOMs}
\subsection{Details of the algorithm}
 We focus on a system with $L$ sites, which is described by the~Hilbert space with dimension~$Z$ spanned by energy eigenstates $H|n\rangle=E_n|n\rangle$. We consider a set of~$D_{O}$ observables $\{A^1,...,A^{D_O} \}$ and denote their matrix elements as $A^{s}_{nm}=\langle n | A^{s} |m \rangle$. 
The latter operators are local, traceless and orthonormal with respect to the Hilbert-Schmidt (HS) product introduced in Eq.~(\ref{hs}) in the~main text, i.e.,
\begin{equation}
\langle A^s A^{s'} \rangle=
\frac{1}{Z}{\rm Tr}(A^s A^{s'}) =\delta_{ss'}. \label{suport1}
\end{equation}

Nonergodic systems retain memory of the~initial state, so that the correlation function $ \lim_{t\to \infty} \lim_{L\to \infty}  \langle A^s(t) A^s \rangle $ is nonzero for the majority of $A^{s}$. This property leads to the nonvanishing stiffness
\begin{equation}
\lim_{\tau\rightarrow\infty} \frac{1}{\tau}\int_{0}^{\tau} {\rm d}t \langle A^s(t) A^s \rangle
=\langle \bar{A}^s A^s \rangle = \langle \bar{A}^s \bar{A}^s \rangle=|| \bar{A}^s ||^2 , \label{stiff}
\end{equation}
which equals the squared HS norm of the time-averaged operator, $\bar{A}^s= \sum_{E_n=E_m}A^{s}_{nm} |n\rangle\langle m|$, see Eq.~(\ref{ta}) in the~main text. For clarity, we present explicit expressions only for the case with no degeneracies in the many-body spectrum, when
 \mbox{$\bar{A}^s= \sum_{n=1}^Z A^{s}_{nn} |n\rangle\langle n|$}.

The nonergodic behavior of operators $A^s$ is encoded in the matrix elements of their time-averaged counterparts
$\bar{A}^{s}$. Therefore, we build the matrix
\begin{equation}
    \mathcal{R}=
\begin{bmatrix}
A_{11}^{1} & A_{11}^{2} & . & . & . & A_{11}^{D_O}\\ 
A_{22}^{1} & A_{22}^{2} & . & . & . & A_{22}^{D_O}\\ 
. & . & . & . & . & .\\
. & . & . & . & . & .\\
. & . & . & . & . & .\\
A_{ZZ}^{1} & A_{ZZ}^{2} & . & . & . & A_{ZZ}^{D_O} \label{suprdef}
\end{bmatrix}\;,
\end{equation}
which $s$th column gathers all matrix elements of $\bar{A}^{s}$.
When degeneracies are present in the many-body spectrum then each row corresponds to a~different pair of degenerated energy eigenstates and the~number of rows is larger than $Z$. We also assume that $D_O<Z$, so that the rank of $\mathcal{R}$ is not larger than $D_O$. 
Next, we perform the singular value decomposition ($\mathcal{R}=\mathcal{U} \tilde{\Lambda}\mathcal{V}^{T} $),
\begin{equation}
    \mathcal{R}_{ns}=A^s_{nn}= {\rm Tr}\left(|n\rangle \langle n| A^s\right)=\sum_{\beta=1}^{D_O} \mathcal{U}_{n \beta} \tilde{\lambda}_{\beta} ( \mathcal{V}^{T})_{\beta s } \;, \label{supsvd1}
\end{equation}
where the number of non-zero singular values $\tilde{\lambda}_\alpha$  is given by the rank of $\mathcal{R}$. We take positive $\tilde{\lambda}_{\beta}$, which are sorted in descending order. Using the orthogonality of  matrices $ \mathcal{U}$ and $\mathcal{V}$, we solve Eq.~(\ref{supsvd1}) for the singular values
\begin{equation}
\tilde{\lambda}_{\beta} \delta_{\beta \beta'}= {\rm Tr}\left[ \left( \sum_{n=1}^{Z}   \mathcal{U}_{n \beta'} |n\rangle \langle n|   \right)   \left( \sum_{s=1}^{D_O} \mathcal{V}_{s \beta} A^s \right) \right], 
\end{equation}
and rewrite the above equation in terms of the Hilbert Schmidt inner product
\begin{eqnarray}
\left< Q^{\beta'}  B^{\beta}  \right> &=& \frac{\tilde{\lambda}_{\beta}}{\sqrt{Z}}  \delta_{\beta \beta'} \equiv \sqrt{\lambda_{\beta}}  \delta_{\beta \beta'}  \;, \label{supcsne} \\
Q^{\beta}&=& \sqrt{Z} \sum_{n=1}^{Z}   \mathcal{U}_{n \beta} |n\rangle \langle n|  \;,  \label{supqdef} \\ 
B^{\beta}&=&\sum_{s=1}^{D_O} \mathcal{V}_{s\beta}A^{s}\;, \label{supbdef}
\end{eqnarray} 
cf. Eq.~(\ref{suport1}).
We have introduced the factor $\sqrt{Z}$ in Eq.~(\ref{supqdef}) to ensure that the operators $Q^{\beta}$ are orthonormal, i.e., $\langle Q^{\beta}  Q^{\beta'} \rangle= \delta_{\beta \beta'}$. We note that the orthogonal transformation of $A^{s}$ in Eq.~(\ref{supbdef}) preserves their orthogonality and normalization, i.e., $\langle B^{\beta}  B^{\beta'} \rangle= \delta_{\beta \beta'}$. 
For clarity, we explicitly account for the norms of operators which appear in Eq.~(\ref{supcsne}) 
 \begin{equation}
 \left< Q^{\beta'}  B^{\beta}  \right> =\delta_{\beta \beta'} \sqrt{\lambda_{\beta}} \;  ||Q^{\beta'} || \; ||  B^{\beta} || \; .\label{supcsne1}
 \end{equation}
Comparing  the above equation with  the Cauchy Schwartz inequality one gets $ \lambda_{\beta} \le 1$ and, most importantly,  for $\lambda_{\beta}=1$
one obtains $Q^{\beta}=B^{\beta}$. All $Q^{\beta}$ are conserved by construction, as they are expressed via projectors on eigenstates of the Hamiltonian.
We also note that all $B^{\beta}$ represent local operators, as they are expressed as linear combinations of local operators $A^s$. Therefore for 
$\lambda_{\beta}=1$, the operator $Q^{\beta}=B^{\beta}$ is both local and conserved, hence it is a LIOM. 

While we do not study quasilocal integrals of motion in the present work, we note that they can also be singled out by this procedure. Such operators are not strictly local,
but their projections on certain local operators do not vanish in the thermodynamic limit ($ L \to \infty$). From Eq.~(\ref{supcsne1}) one finds that the quasilocal integrals of motion are represented by $Q_{\beta}$ for which $0<\lim_{L \to \infty} \lambda_{\beta} <1 $. 

\subsection{Relation to the data compression problem and the Mazur bound}

 The  elements of ${\cal R}$ in Eq.~(\ref{suprdef}) store information about the nonergodic behavior of the studied set of operators, $\{A^1,...,A^{D_O} \}$.
In general,  the rank of ${\cal R}$ equals $D_O$. One may  pose a question about a possibility of compressing the data stored in ${\cal R}$.
Formally, one looks for a matrix  ${\cal R^{\parallel}}$ of a fixed rank $D_L \ll D_O$, which minimizes the Hilbert Schmidt norm $|| {\cal R} - {\cal R^{\parallel}} ||$. The solution of this variational problem is given by the~Eckart–Young–Mirsky theorem. Namely, the matrix elements of  ${\cal R^{\parallel}}$
are given by the same singular value decomposition that we have used for identification of LIOMs,
 \begin{equation}
 \mathcal{R^{\parallel}}_{ns}=\sum_{\beta=1}^{D_L} \mathcal{U}_{n \beta} \tilde{\lambda}_{\beta} ( \mathcal{V}^{T})_{\beta s } \;. \label{supsvd2}
\end{equation}
 In contrast to Eq.~(\ref{supsvd1}), the summation in Eq.~(\ref{supsvd2}) is carried out only over $D_L$ largest singular values. 
 Below, we demonstrate that the data compression via the~Eckart–Young–Mirsky theorem is equivalent to the procedure projecting the studied operators onto LIOMs.

 Using Eq.~(\ref{supsvd1}), we express the time-averaged operators $\bar{A}^s$ via the conserved operators introduced in Eq.~(\ref{supqdef}) 
\begin{eqnarray} 
\bar{A}^s & =& \sum_{n=1}^{Z}    A^s_{nn}   |n\rangle \langle n| =\sum_{\beta=1}^{D_O} \left( \sum_{n=1}^{Z}  \mathcal{U}_{n \beta}  |n\rangle \langle n| \right) \tilde{\lambda}_{\beta}  ( \mathcal{V}^{T})_{\beta s } \nonumber \\
&=& \sum_{\beta=1}^{D_O} Q^{\beta} \sqrt{\lambda_{\beta}} ( \mathcal{V}^{T})_{\beta s } \label{suppro3} \\
&= &\sum_{\beta=1}^{D_O} Q^{\beta}  \langle Q^{\beta} \bar{A}^s. \rangle\; 
  \label{suppro1}
\end{eqnarray}   
In the last step, we have used the relation \mbox{$\langle Q^{\beta} \bar{A}^s \rangle=\sqrt{\lambda_{\beta}} ( \mathcal{V}^{T})_{\beta s }$}, which can be established from $\bar{A}^s$ given by Eq.~(\ref{suppro3}) and the orthogonality condition $\langle Q^{\beta} Q^{\beta'} \rangle=\delta_{\beta \beta'}$.
Finally, the identity relation for the time-averaged operators
\mbox{$ \langle Q^{\beta} \bar{A}^s \rangle=\langle \bar{Q}^{\beta}  A^s \rangle=\langle Q^{\beta} A^s \rangle $} allows us to rewrite Eq.~(\ref{suppro1}) in the 
following form
\begin{equation}
\bar{A}^s= \sum_{\beta=1}^{D_O}   \langle Q^{\beta} A^s \rangle Q^{\beta}\;. \label{suppro4}
\end{equation}

One may repeat the reasoning from Eqs. (\ref{suppro3})-(\ref{suppro4}) taking into account only $D_L$ largest singular values. Within such modified procedure we obtain the following operators 
 \begin{equation}
\bar{A}^{s \parallel}=  \sum_{\beta=1}^{D_L} Q^{\beta} \sqrt{\lambda_{\beta}} ( \mathcal{V}^{T})_{\beta s } =\sum_{\beta=1}^{D_L}   \langle Q^{\beta} A^s \rangle Q^{\beta}, \label{suppro5}
 \end{equation}
 whose matrix elements are stored in the compressed matrix ${\cal R}^{\parallel}$, i.e., \mbox{ $\langle n| \bar{A}^{s \parallel} |n\rangle =  \mathcal{R^{\parallel}}_{ns}$}. 
 Therefore, the solution of the data compression problem for the matrix $ {\cal R}$  is equivalent to the projection of the studied operators on
  LIOMS with $\lambda_{\beta}=1$ and (depending on the choice of $D_L$)  on other conserved operators with the  largest projections ($\lambda_{\beta}$), see   Eq.~(\ref{supcsne}).

The Mazur bound~\cite{MAZUR1969533,SUZUKI1971277} for the stiffnesses introduced in Eq.~(\ref{stiff}) can be easily obtained from Eq.~(\ref{suppro4}) together with the orthogonality condition $\langle Q^{\beta} Q^{\beta'} \rangle=\delta_{\beta \beta'}$,
\begin{eqnarray}
  \langle \bar{A}^s \bar{A}^s \rangle &=&  \sum_{\beta=1}^{D_O}   \langle Q^{\beta} A^s \rangle^2 \ge
  \sum_{\beta=1}^{D_L}   \langle Q^{\beta} A^s \rangle^2.
\end{eqnarray}
Finally, we note that the stiffness of operators ${B}^{\beta}$ defined in Eq.~(\ref{supbdef}) has a particularly simple form. Using
 Eqs. (\ref{suppro4}) and (\ref{supcsne}), one arrives at
\begin{eqnarray}
 \bar{B}^{\beta}&=&\sum_{s=1}^{D_O} \mathcal{V}_{s\beta}\bar{A}^{s}  = 
\sum_{s=1}^{D_O} \mathcal{V}_{s\beta} \sum_{\beta'=1}^{D_O}   \langle Q^{\beta'} A^s   \rangle Q^{\beta'} \nonumber \\
&=& \sum_{\beta'=1}^{D_O}   \langle Q^{\beta'} \left( \sum_{s=1}^{D_O} \mathcal{V}_{s\beta} A^s \right)   \rangle Q^{\beta'} \nonumber \\
&=&   \sum_{\beta'=1}^{D_O}   \langle Q^{\beta'} B^{\beta}   \rangle Q^{\beta'}=\sqrt{\lambda_{\beta}} Q^{\beta}.
\end{eqnarray}
The Hilbert-Schmidt norm is, thus, given by $|| \bar{B}^{\beta}||^2=\lambda_{\beta}$. If $\lambda_{\beta}=1$, then the operator $B^{\beta}$ has no off-diagonal matrix elements that are be eliminated by the time-averaging and \mbox{ $|| \bar{B}^{\beta}||=|| B^{\beta}||=1$}. Therefore, we reach the same conclusion as from Eq.~(\ref{supcsne}), i.e., if $\lambda_{\beta}=1$ then the operator $B^{\beta}$
is a LIOM.

The most demanding part of numerical implementation of the above algorithm concerns
the construction of the matrix ${\cal R}$. In order to single out the matrix elements of time-averaged operators one needs to carry out diagonalization of the studied Hamiltonian. In the case of spin ($s=1/2$) systems (or spinless fermions) one is typically restricted to systems containing $L \sim 20$ sites.   

\section{Orthogonalization of operators}

In order to orthogonalize the set of operators $O^{j}$ with $j\in\{1,...,D_O\}$, it is necessary to construct their traceless counterparts
\begin{equation}
    \underline{O}^{j}=O^{j}-c(j)\;,
\end{equation} 
where $c(j)$ is a real constant fixed by the condition $\text{Tr}(\underline{O}^{j})=0$. Next, we build the matrix of the Hilbert-Schmidt products
\begin{equation}
    R=
\begin{bmatrix}
\langle \underline{O}^{1} \underline{O}^{1} \rangle & \langle \underline{O}^{1} \underline{O}^{2} \rangle & . & . & . & \langle \underline{O}^{1} \underline{O}^{D} \rangle\\ 
\langle \underline{O}^{2} \underline{O}^{1} \rangle & \langle \underline{O}^{2} \underline{O}^{2} \rangle & . & . & . & \langle \underline{O}^{2} \underline{O}^{D} \rangle\\ 
. & . & . & . & . & .\\
. & . & . & . & . & .\\
. & . & . & . & . & .\\
\langle \underline{O}^{D} \underline{O}^{1} \rangle & \langle \underline{O}^{D} \underline{O}^{2} \rangle & . & . & . & \langle \underline{O}^{D} \underline{O}^{D} \rangle
\end{bmatrix}\;,
\end{equation}
which is real and symmetric, and we solve the eigenproblem
\begin{equation}
    R=U\mathcal{D}U^T,
\end{equation}
where $\mathcal{D}$
is a diagonal matrix, $ (\mathcal{D})_{ii'}=\sigma_{i}\delta_{ii'}$, with dimension $D_O\times D_O$ and positive eigenvalues $\sigma_{i}$. The matrix $U$ with dimension $D_O\times D_O$ determines the orthogonal transformation of operators, so that
\begin{equation}
    A^{i}=\frac{1}{\sqrt{\sigma_i}}\sum_{j=1}^{D_O} {U}_{ji}\underline{O}^{j} 
\end{equation}
are orthogonal and normalized
\begin{equation}
\begin{split}
    \langle A^{i} A^{i'} \rangle & = \frac{1}{\sqrt{\sigma_i}\sqrt{\sigma_{i'}}}\sum_{j,j'=1}^{D_O} {U}_{ji}\langle \underline{O}^{j} \underline{O}^{j'} \rangle {U}_{j'i'}\\
    & = \frac{\sigma_{i'}}{\sqrt{\sigma_i}\sqrt{\sigma_{i'}}}\sum_{j}^{D_O} {U}_{ji} {U}_{ji'}=\delta_{ii'}\;.
\end{split}
\end{equation}
We emphasize that if $O^j$ are local then $A^{i}$ inherit this property.

\section{Spectral statistics}

A popular measure of quantum chaos is the statistics of the spacings $\delta_i=E_{i}-E_{i-1}$ between the nearest energy levels $E_{i}$ and $E_{i-1}$. In order to avoid the spectral unfolding, it is convenient to study the ratio
\begin{equation}
    \tilde{r}_{i}=\frac{\text{min}(\delta_{i+1},\delta_{i})}{\text{max}(\delta_{i+1},\delta_{i})},
\end{equation}
which is restricted to the interval $[0,1]$. Energy levels within each symmetry sector of ergodic systems are correlated and experience level repulsion, so that the distribution of ratios $\tilde{r}$ agrees with the prediction of the Gaussian orthogonal ensemble,
\begin{equation}
\label{eqPoisson}
    P_\text{GOE}(\tilde{r})=\frac{27}{4}\frac{\tilde{r}+\tilde{r}^2}{(1+\tilde{r}+\tilde{r}^2)^{5/2}},
\end{equation}
which vanishes in the limit $\tilde{r}\rightarrow 0$, while its mean is $\langle\tilde{r}\rangle_\text{GOE}\approx 0.536$~\cite{PhysRevLett.110.084101,PhysRevB.75.155111}.
Simultaneously, energy levels of nonergodic systems, characterized by an extensive number of LIOMS, can be massively degenerated. If this is not the case, then energy levels are uncorrelated and the distribution of ratios $\tilde{r}$ agrees with the Poisson distribution,
\begin{equation}
\label{eqPoisson}
    P_\text{P}(\tilde{r})=\frac{2}{(1+\tilde{r})^2},
\end{equation}
which is nonzero in the limit $\tilde{r}\rightarrow 0$, while its mean is $\langle\tilde{r}\rangle_\text{P}\approx 0.386$~\cite{PhysRevLett.110.084101,PhysRevB.75.155111}.

\begin{figure}[t!]
\centering
\includegraphics[width=0.85\columnwidth]{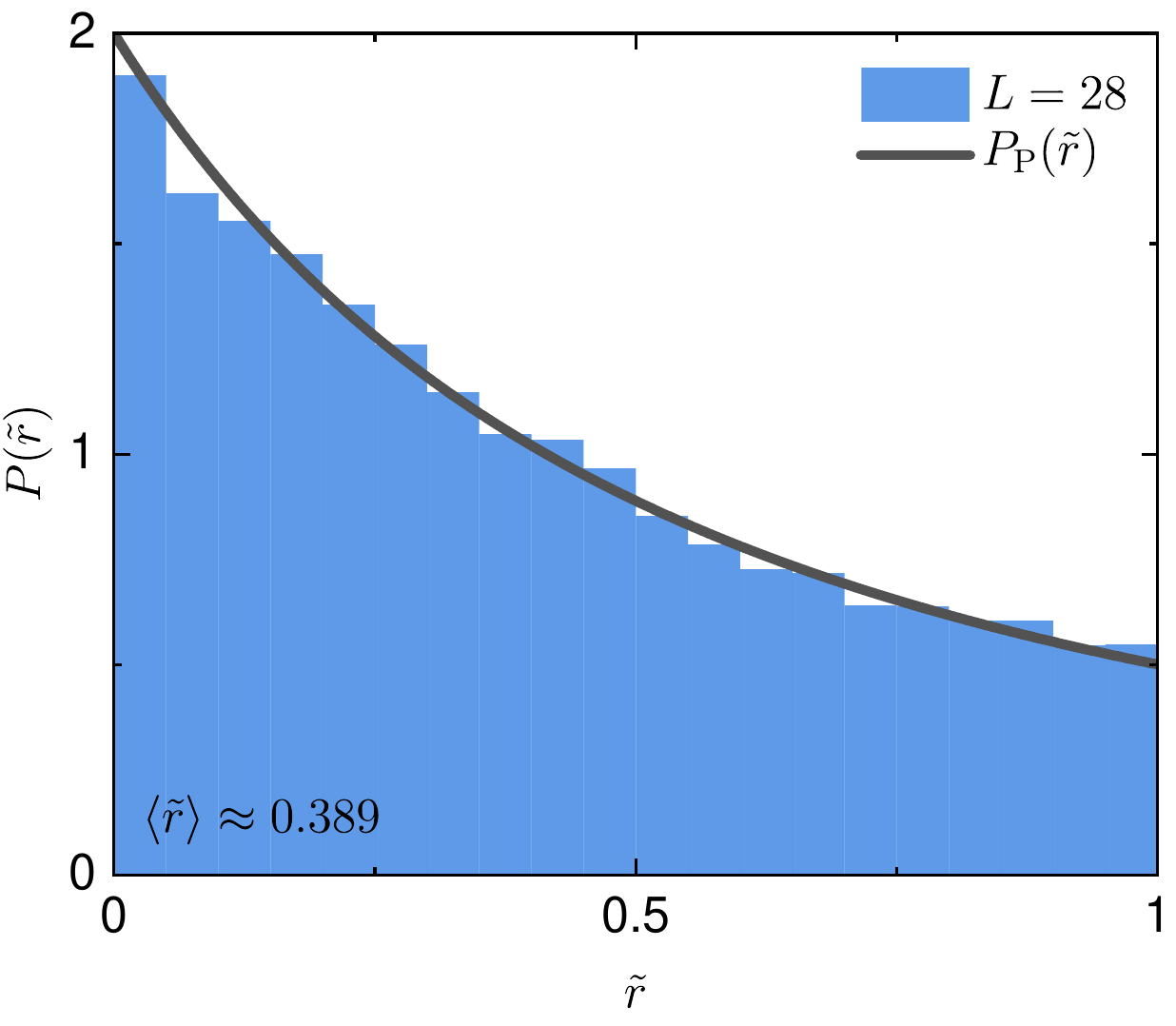}
\caption{The spectral statistics of the PH model from Eq.~(\ref{eqSpectral}) with $L=28$ sites. We consider the largest symmetry sector with a fixed dipole moment $M=\frac{L(L+1)}{4}=203$ and sublattice particle number $n_\text{even}=\frac{L}{4}=7$, which dimension is $\tilde{Z}=417540$. The histogram has been calculated out of $2000$ energy eigenstates in the middle of the spectrum and averaged over $20$ Hamiltonian realizations. The solid line corresponds to the Poisson distribution from Eq.~(\ref{eqPoisson}).}
\label{figS2}
\end{figure}

We have verified that there are many degeneracies in the spectrum of the PH model (not shown). 
Following Ref.~\cite{Moudgalya_2021} and \cite{PhysRevB.103.134207}, we lift these degeneracies by adding a minor disorder to the pair hopping and onsite potential, which preserves the Krylov structure of the Hilbert space. Specifically, we consider a modified Hamiltonian, which can be written as
\begin{equation}
\label{eqSpectral}
    \tilde{H}_1=\sum_{i=1}^{L-3} t_{i}(c^\dagger_{i}c^\dagger_{i+3}c_{i+2}c_{i+1} + \text{H.c.})+\sum_{i=1}^{L}\epsilon_{i}c^\dagger_{i}c_{i}\;,
\end{equation}
where $t_i$ and $\epsilon_i$ are identically distributed random numbers drawn from intervals $[0.99,1.01]$ and $[-0.01,0.01]$, respectively. In Fig.~\ref{figS2}, we present the spectral statistics of $\tilde{H}_1$ from Eq.~(\ref{eqSpectral}) with $L=28$ sites. We consider the largest symmetry sector with the dipole moment $M=\frac{1}{Z}\text{Tr}(M)=\frac{L(L+1)}{4}=203$ and sublattice particle number $n_\text{even}=\frac{1}{Z}\text{Tr}(n_\text{even})=\frac{L}{4}=7$, which dimension is $\tilde{Z}=417540$. The histogram of ratios $\tilde{r}$ has been evaluated in the middle of the spectrum from $2000$ energy eigenstates and averaged over $20$ realizations of
$\{t_i\}$ and $\{\epsilon_{i} \}$. We observe a remarkable agreement with the Poisson distribution from Eq.~(\ref{eqPoisson}) with the mean ratio $\langle\tilde{r}\rangle\approx 0.389$. This is a clear evidence that the PH model from Eq.~(\ref{eqSpectral}) supports additional integrals of motion beyond $M$ and $n_\text{even}$, which are not fixed in the studied symmetry sector \cite{doi:10.1143/JPSJ.74.1992}. The latter observation goes in line with the general conclusion of the main text concerning the presence of LIOMs.

\section{Hamiltonian from Eq.~(\ref{eqH2}) without the last term}

\begin{figure}[t!]
\centering
\includegraphics[width=\columnwidth]{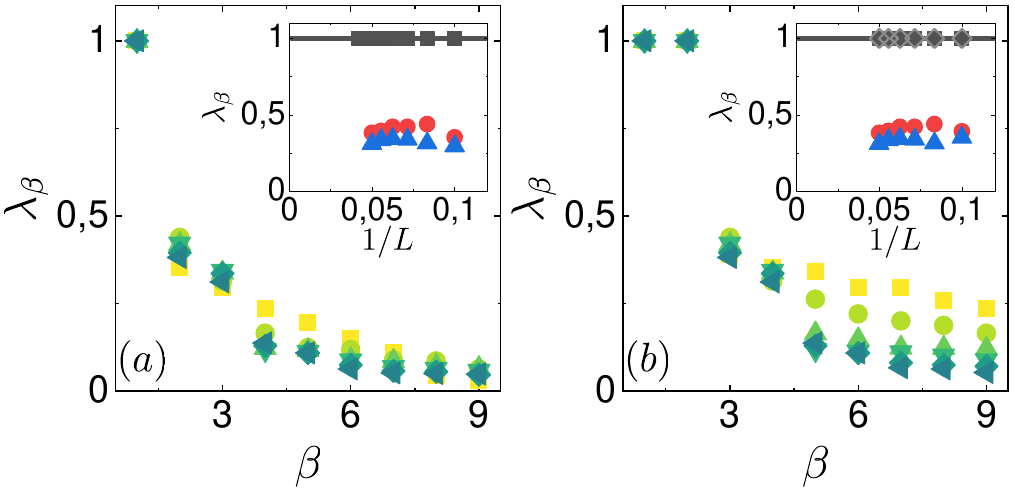}
\caption{The norms $\lambda_\beta$ for $\tilde{H}_{2}$ from Eq.~(\ref{Happ}), which is the~simplest extension of the PH model. Results are presented for $\beta\le 9$, $L=10,...,20$ and various sets of operators: (a)~$A^i \in {\cal N}$ and (b)~$A^i \in {\cal \tilde{N}}_{2E}$. The insets show the finite size scalings of the largest $\lambda_\beta$.}
\label{figS1}
\end{figure}

The simplest extension of the PH model is
\begin{equation}
\label{Happ}
    \tilde{H}_{2}=H_{1}+\sum_{i=1}^{L-4} (c^\dagger_{i}c^\dagger_{i+4}c_{i+3}c_{i+1} + \text{H.c.})\;,
\end{equation}
where $H_1$ corresponds to the Hamiltonian from Eq.~(\ref{eqH1}) in the main text. For this model, we have calculated the~HS norms $\lambda_\beta$, where $B^\beta$ are linear combinations of $A^{i}\in\mathcal{N}$, which are site occupations. The results of numerical calculations are presented in Fig.~\ref{figS1}(a). Only one LIOM ($\lambda_{1}=1$) can be observed and it represents the dipole moment $M$ (not shown). Moreover, we have considered an extended set $\tilde{\mathcal{N}}_{2E}$, which includes all operators from $\mathcal{N}$ as well as the pair hopping terms $c^\dagger_{i}c^\dagger_{i+3}c_{i+2}c_{i+1} + \text{H.c.}$ and $c^\dagger_{i}c^\dagger_{i+4}c_{i+3}c_{i+1} + \text{H.c.}$ This allows to construct $\tilde{H}_{2}$ as a linear combination of $A^{i}\in\tilde{\mathcal{N}}_{2E}$, see Fig.~\ref{figS1}(b). In this case, one obtains two LIOMs that represent the Hamiltonian and the dipole moment. However, the available data do not allow to rule out the possibility that for other  
$\lambda_\beta$ one gets $0<\lim_{L\to \infty} \lambda_\beta <1$. The latter may be interpreted as a presence of additional integrals
of motion, which cannot be entirely expressed as linear combinations of operators from the studied set but have nonvanishing projections on operators belonging to this set. Such conserved quantities do not need to be local, e.g., they can be quasilocal in a sense of Ref.~\cite{PhysRevLett.106.217206}.

\section{Derivation of effective models of tilted chain}

The derivation of the PH model, Eq.~(\ref{eqH1}), has been performed so far by the Schrieffer-Wolff transformation of the original Stark model, Eq.~(\ref{eqH3}), assuming large fields $F \gg t$. Specifically, the nearest-neighbor hopping term has been eliminated and the lowest order terms in the~expansion in the strength of $t/F$ have been kept~\cite{Moudgalya_2021}. The~effective model conserves by construction the dipole moment $M$. In general, such procedure can be extended to establish higher order terms, as the ones in the EPH model, Eq.~(\ref{eqH2}). However, one can  choose an alternative way and assume weak interactions  $V \lesssim F,t$. Then, one can rewrite the Hamiltonian, Eq.~(\ref{eqH3}), within the basis of non-interacting Stark single-particle states, i.e.,
\begin{equation}
H_0 = \sum_j \epsilon_j  a^\dagger_j a_j, \qquad c^\dagger_i = \sum_j \alpha_{ij} a^\dagger_j, 
\end{equation}
 where (neglecting the boundary effects) energies are equidistant $\epsilon_{j+1}= \epsilon_j
 +F$ and wave functions are localized $\alpha_{ij}= f(j-i)$ in a range $|j-i| \lesssim t/F$. The~interaction term in this basis becomes
 \begin{eqnarray}
H' &=& V  \sum_i n_i n_{i+1} = 
 V \sum_{jklm} \chi_{jk}^{lm} ~a^\dagger_l a^\dagger_m a_k a_j, \\
\chi^{lm}_{jk} &=& \sum_i \alpha_{il} \alpha_{i+1,m} \alpha_{i+1,k} \alpha_{ij}. 
\end{eqnarray} 
Such an approach has been previously used in the analysis of an analogous MBL problem~\cite{prelovsek18}, where the basis of Anderson single-particle states is the~relevant one.

One can classify the terms in $H'$ by the number of sites involved, i.e., $H'_{2}$, $H'_{3}$, $H'_{4}$. Note that
\begin{equation}
H^\prime_2 = 2 V \sum_{j<k} ( \chi_{jk}^{jk} - \chi_{jk}^{kj} ) n_j n_k,
\end{equation}
is the Hartree-Fock term conserving $M$. The most important terms among the remaining ones are those that also conserve $M$, in particular those that emerge in $H_4'$,
\begin{equation}
H^\prime_{4dr} = \sum_j \zeta_{dr}[a_{j-r}^\dagger a_j a_{j+d+r}^\dagger a_{j+d} + \mathrm{H.c.}],
\end{equation}
which generate pair hoppings of different range $r \geq 1$ and distance $d \geq 1$.
Keeping only $\tilde H'_4 = \sum_{d,r \geq 1} H^\prime_{4dr}$ results in the model containing terms of infinite range. Still, $\zeta_{dr}$ decay
with $d,r \gg 1$ for large fields $F\gtrsim V$. This allows constructing extensions of the PH model, as the~ones from Eqs.~(\ref{Happ}) and (\ref{eqH2}). In principle, one could consider also remaining 
$H'_4 \neq  \tilde H'_4$ and  $H'_3$ and eliminate them by an appropriate Schrieffer-Wolff transformation. However, the generated terms would be of a higher order in the interaction, $V^m$ with $m\geq 2$. 

\section{Subdiffusion from dynamical structure factor}

The transport (diffusion) properties of the selected models at $T \to \infty$ can be studied  via the dynamical structure factor (the density correlation function) 
\begin{equation}
    S(q,\omega)= \frac{1}{\pi} \mathrm{Re} \int_0^\infty dt \mathrm{e}
^{i \omega t}   \langle n_{q} (t) n_{-q} \rangle,
\end{equation}
with
\begin{equation}
    n_q=\frac{1}{\sqrt{L}} \sum_{i} \cos(q(i-L/2))n_i.
\end{equation}
 The~particle number is conserved, so the dynamical structure factor in the low-$q,\omega$ regime should take the hydrodynamic form (see, e.g., \cite{herbrych12})
\begin{equation}
S(q,\omega) \sim \frac{1}{\pi}  \frac{\chi^0_q \Gamma_q}{\omega^2+\Gamma^2_q}, \qquad \chi^0_q =\int d\omega  S(q,\omega),
\end{equation}
with the relaxation rate $\Gamma_q \sim D q^z$ where $z>0$, the~corresponding susceptibility $\chi^0_q = \langle n_q n_{-q} \rangle
\sim \chi^0 \sim (1/2) \bar n (1- \bar n)$ and the particle density $\bar n =N/L$.
 
Results for $S(q,\omega)$ in the EPH model, Eq.~(\ref{eqH2}), were obtained with the micro-canonical Lanczos method 
(MCLM) \cite{long03,prelovsek13}.
Since the EPH model conserves both the particle number $N$ and the dipole moment $M$, we performed calculations in the largest symmetry sector with $N=L/2$ and $M=0$,
reaching the system size $L=32$ with the~Hilbert space dimension $N_{st} \sim 10^7$. To obtain high-enough
frequency resolution $\delta \omega \sim 4.10^{-4}$, we had to use a large number
of Lanczos steps $N_L \sim 5.10^4$. 

In Fig.~\ref{fig2}(c), we present
results for $S(q,\omega)$
for two lowest nonzero $q=2 m\pi/L$ with $m=1,2$. The main conclusion is that the numerical results confirm the subdiffusive scaling with $z=4$ in the low-$q,\omega$ regime, since 
$\pi S(q,\omega \sim 0)) = \chi^0 /\Gamma_q \sim \chi^0 /(Dq^4)$.

\end{document}